\documentclass[aps,prd,groupedaddress,showpacs]{revtex4}
\usepackage{graphicx,epsf,amssymb,amsmath,latexsym}
\newcommand{\be}{\begin{eqnarray}}
\newcommand{\ee}{\end{eqnarray}}
\def\ben{\begin{equation}}
\def\een{\end{equation}}
\def\bena{\begin{eqnarray}}
\def\eena{\end{eqnarray}}

\begin{document}







%
\title{Information Loss in Black Holes}

\vspace{.3in}

\author{S.W.Hawking}
\email{...}
\affiliation{DAMTP, Center for Mathematical Sciences, university of Cambridge, Wilberforce Road, Cambridge CB3 0WA, UK }

\vskip.3in
\begin{abstract}
%

The question of whether information is lost in black holes is investigated using Euclidean path integrals. The formation and evaporation of black holes is regarded as a scattering problem with all measurements being made at infinity. This seems to be well formulated only in asymptotically AdS spacetimes. The path integral over metrics with trivial topology is unitary and information preserving. On the other hand, the path integral over metrics with non-trivial topologies leads to correlation functions that decay to zero. Thus at late times only the unitary information preserving path integrals over trivial topologies will contribute. Elementary quantum gravity interactions do not lose information or quantum coherence.    
\end{abstract}
\maketitle
%

\setcounter{equation}{0}
\section{Introduction}

The black hole information paradox started in 1967 when Werner Israel showed that the Schwarzschild metric was the only static vacuum black hole solution \cite{Israel}. This was then generalized to the no hair theorem, the only stationary rotating black hole solutions of the Einstein Maxwell equations are the Kerr Newman metrics \cite{NoHair}. The no hair theorem implied that all information about the collapsing body was lost from the outside region apart from three conserved quantities: the mass, the angular momentum, and the electric charge.

This loss of information wasn't a problem in the classical theory. A classical black hole would last for ever and the information could be thought of as preserved inside it, but just not very accessible. However, the situation changed when I discovered that quantum effects would cause a black hole to radiate at a steady rate \cite{Stephen1}. At least in the approximation I was using the radiation from the black hole would be completely thermal and would carry no information\cite{Stephen2}. So what would happen to all that information locked inside a black hole that evaporated away and disappeared completely?  It seemed the only way the information could come out would be if the radiation was not exactly thermal but had subtle correlations. No one has found a mechanism to produce correlations but most physicists believe one must exist. If information were lost in black holes, pure quantum states would decay into mixed states and quantum gravity wouldn't be unitary.

I first raised the question of information loss in 75 and the argument continued for years without any resolution either way. Finally, it was claimed that the issue was settled in favor of conservation of information by ADS-CFT. 
ADS-CFT is a conjectured duality between string theory in anti de Sitter space and a conformal field theory on the boundary of anti de Sitter space at infinity \cite{ADSCFTreviews }. Since the conformal field theory is manifestly unitary the argument is that string theory must be information preserving. Any information that falls in a black hole in anti de Sitter space must come out again. But it still wasn't clear how information could get out of a black hole. It is this question, I will address in this paper. 

\section{Euclidean Quantum Gravity}

Black hole formation and evaporation can be thought of as a scattering process. One sends in particles and radiation from infinity and measures what comes back out to infinity. All measurements are made at infinity, where fields are weak and one never probes the strong field region in the middle. So one can't be sure a black hole forms, no matter how certain it might be in classical theory. I shall show that this possibility allows information to be preserved and to be returned to infinity. 

I adopt the Euclidean approach \cite{EQG}, the only sane way to do quantum gravity nonperturbatively. One might think one should calculate the time evolution of the initial state by doing a path integral over all positive definite metrics that go between two surfaces that are a distance $T$ apart at infinity. One would then Wick rotate the time interval $T$ to the Lorentzian. 

The trouble with this is that the quantum state for the gravitational field on an initial or final space-like surface is described by a wave function which is a functional of the geometries of space-like surfaces and the matter fields 

\begin{equation}
\Psi[h_{ij}, \phi, t]
\end{equation}
where $h_{ij}$ is the three metric of the surface, $\phi$ stands for the matter fields and $t$ is the time at infinity. 
However there is no gauge invariant way in which one can specify the time position of the surface in the interior. This means one can not give the initial wave function without already knowing the entire time evolution. 

One can measure the weak gravitational fields on a time like tube around the system but not on the caps at top and bottom which go through the interior of the system where the fields may be strong. One way of getting rid of the difficulties of caps would be to join the final surface back to the initial surface and integrate over all spatial geometries of the join.   If this was an identification under a Lorentzian time interval $T$ at infinity, it would introduce closed time like curves. But if the interval at infinity is the Euclidean  distance $\beta$ the path integral gives the partition function for gravity at temperature $\Theta=\beta^{-1}$.  

\begin{eqnarray}
Z(\beta)&=&\int DgD\phi e^{-I[g, \phi]}\nonumber \\
&=&\text{Tr}(e^{-\beta H})
\end{eqnarray}

There is an infrared problem with this idea for asymptotically flat space.  The partition function is infinite because the volume of space is infinite. This problem can be solved by adding a small negative cosmological constant $\Lambda$ which makes the effective volume of the space the order of $\Lambda^{-3/2}$. It will not affect the evaporation of a small black hole but it will change infinity to anti-de Sitter space and make the thermal partition function finite.

It seems that asymptotically anti-de Sitter space is the only arena in which particle scattering in quantum gravity is well formulated. Particle scattering in asymptotically flat space would involve null infinity and Lorentzian metrics, but there are problems with non-zero mass fields, horizons and singularities. Because measurements can be made only at spatial infinity, one can never be sure if a black hole is present or not.

\section{ The Path Integral }

The boundary at infinity has topology $S^1 \times S^2$. The path integral that gives the partition function is taken over metrics of all topologies that fit inside this boundary. The simplest topology is the trivial topology $S^1 \times D^3$ where $D^3$ is the three disk. The next simplest topology and the first non-trivial topology is $S^2 \times D^2$. This is the topology of the Schwarzschild anti-de Sitter metric. There are other possible topologies that fit inside the boundary but these two are the important cases, topologically trivial metrics and the black hole. The black hole is eternal: it can not become topologically trivial at late times.

The trivial topology can be foliated by a family of surfaces of constant time. The path integral over all metrics with trivial topology can be treated canonically by time slicing. The argument is the same as for the path integral for ordinary quantum fields in flat space. One divides the time interval $T$ into time steps $\Delta t$. In each time step one makes a linear interpolation of the fields $q_i$ and their conjugate momenta between their values on succesive time steps. This method applies equally well to topologically trivial quantum gravity and shows that the time evolution (including gravity) will be generated by 
a Hamiltonian. This will give a unitary mapping between quantum states on surfaces separated by a time interval $T$ at infinity. 

This argument can not be applied to the non-trivial black hole topologies. They can not be foliated by a family of surfaces of constant time because they don't have any spatial cross-sections that are a three cycle, modulo the boundary at infinity. Any global symmetry would lead to conserved global charges on such a three cycle. These would prevent correlation functions from decaying in topologically trivial metrics. Indeed, one can regard the unitary Hamiltonian evolution of a topologically trivial metric as a global conservation of information flowing through a three cycle under a global time translation. On the other hand, non-trivial black hole topologies won't have any conserved quantity that will prevent correlation functions from decaying. It is therefore very plausible that the path integral over a topologically non trivial metric gives correlation functions that decay to zero at late Lorentzian times. This is  born out by explicit calculations. The correlation functions decay as more and more of the wave falls through the horizon into the black hole.   

\section{Giant Black Holes }

In a thought provoking paper \cite{Maldacena1}, Maldacena considered how the loss of information into black holes in AdS could be reconciled with the unitarity of the CFT on the boundary of AdS. He studied the canonical ensemble for AdS at temperature $\beta^{-1}$. This is given by the path integral over all metrics that fit inside the boundary $S^1\otimes S^2$ where the radius of the $S^1$ is $\beta$ times the radius of the $S^2$. For $\beta <<  \Lambda$ there are three classical solutions that fit inside the boundary: periodically identified AdS, a small black hole and a giant black hole. If one normalizes AdS to have zero action, small black holes have positive action and giant black holes have very large negative action. They therefore dominate the canonical ensemble but the other solutions are important. 

Maldacena considered two point correlation functions in the CFT on the boundary of AdS. The vacuum expectation value $<O(x) O (y)>$ can be thought of as the response at y to disturbances at x corresponding to the insertion of the operator $O$.  It would be difficult to compute in a strongly coupled CFT but by AdS-CFT it is given by boundary to boundary Green functions on the AdS side which can be computed easily. 

The Green functions in the dominant giant black hole solution have the standard form for small separation between $x$ and $y$ but decay exponentially as $y$ goes to late times and most of the effect of the disturbance at $x$ falls through the horizon of the black hole. This looks very like information loss into the black hole. On the CFT side it corresponds to screening of the correlation function whereby the memory of the disturbance at $x$ is washed out by repeated scattering. 

However the CFT is unitary, so theoretically it must be possible to compute its evolution exactly and detect the disturbance at late times from the many point correlation function. All Green functions in the black hole metrics will decay exponentially to zero but Maldacena realized that the Green functions in periodically identified AdS don't decay and have the right order of magnitude to be compatible with unitarity. In this paper I have gone further and shown that the path integral over topologically trivial metrics like periodically identified AdS is unitary.  

So in the end everyone was right in a way. Information is lost in topologically non-trivial metrics like black holes. This corresponds to dissipation in which one loses sight of the exact state. On the other hand, information about the exact state is preserved in topologically trivial metrics. The confusion and paradox arose because people thought classically in terms of a single topology for spacetime. It was either $R^4$ or a black hole. But the Feynman sum over histories allows it to be both at once. One can not tell which topology contributed to the observation, any more than one can tell which slit the electron went through in the two slits experiment. All that observation at infinity can determine is that there is a unitary mapping from initial states to final and that information is not lost. 

\section{Small black holes} 

Giant black holes are stable and won't evaporate away. However, small black holes are unstable and behave like black holes in asymptotically flat space if $M << \Lambda^{-\frac{1}{2}}$ \cite{HawkingPage}. However, in the approach I am using, one can not just set up a small black hole, and watch it evaporate. All one can do, is to consider correlation functions of operators at infinity. One can apply a large number of operators at infinity, weighted with time functions, that in the classical limit would create a spherical ingoing wave from infinity, that in the classical theory would form a small black hole. This would presumably then evaporate away. 

For years, I tried to think of a Euclidean geometry that could represent the formation and evaporation of a single black hole, but without success. I now realize there is no such geometry, only the eternal black hole, and pair creation of black holes, followed by their annihilation. The pair creation case is instructive. The Euclidean geometry can be regarded as a black hole moving on a closed loop, as one would expect. However, the corresponding Lorentzian geometry, represents two black holes that come in from infinity in the infinite past, and accelerate away from each other for ever. The moral of this is that one should not take the Lorentzian analytic continuation of a Euclidean geometry literally as a guide to what an observer would see. Similarly, the formation and evaporation of a small black hole, and the subsequent formation of small black holes from the thermal radiation, should  be represented  by a superposition of trivial metrics and eternal black holes. The probability of observing a small black hole, at a given time, is given by the difference  between the actions. A similar discussion of correlation functions on the boundary shows that the topologically trivial metrics make black hole formation and evaporation unitary and information preserving.
One can restrict to small black holes by integrating the path integral over 
$\beta$ along a contour parallel to the imaginary axis with the factor $e^{\beta E_0}$. This projects  out the states with energy 
$E_0$. 

\begin{equation}
Z(E_0)=\int_{-i\infty}^{+i\infty}d\beta Z(\beta)e^{\beta E_0}
\end{equation}

For $E_0 << \Lambda^{-\frac{1}{2}}$ most of these states will correspond to thermal radiation in AdS which acts like a confining box of volume $\Lambda^{-\frac{3}{2}}$. However, there will be thermal fluctuations which occasionally will be large enough to cause gravitational collapse to form a small black hole. This black hole will evaporate back to thermal AdS. If one now considers correlation functions on the boundary of AdS, one again finds that there is apparent information loss in the small black hole solution but in fact information is preserved by topologically trivial geometries. Another way of seeing that information is preserved in the formation and evaporation of small black holes is that the entropy in the box does not increase steadily with time as it would if information were lost each time a small black hole formed and evaporated.

\section{Conclusions}

In this paper, I have argued that quantum gravity is unitary and information is preserved in black hole formation and evaporation. I assume the evolution is given by a Euclidean path integral over metrics of all topologies. The integral over topologically trivial metrics can be done by dividing the time interval into thin slices and using a linear interpolation to the metric in each slice. The integral over each slice will be unitary and so the whole path integral will be unitary. 

On the other hand, the path integral over topologically non trivial metrics will lose information and will be asymptotically independent of its initial conditions. Thus the total path integral will be unitary and quantum mechanics is safe. 

How does information get out of a black hole? My work with Hartle\cite{HartleHawking} showed the radiation could be thought of as tunnelling out from inside the black hole. It was therefore not unreasonable to suppose that it could carry information out of the black hole.  This explains how a black hole can form and then give out the information about what is inside it while remaining topologically trivial. There is no baby universe branching off, as I once thought. The information remains firmly in our universe. I'm sorry to disappoint science fiction fans, but if information is preserved, there is no possibility of using black holes to travel to other universes. If you jump into a black hole, your mass energy will be returned to our universe but in a mangled form which contains the information about what you were like but in a  state where it can not be easily recognized.  It is like burning an encyclopedia. Information is not lost, if one keeps the smoke and the ashes. But it is difficult to read. In practice, it would be too difficult to re-build a macroscopic object like an encyclopedia that fell inside a black hole from information in the radiation, but the information preserving result is important for microscopic processes involving virtual black holes. If these had not been unitary, there would have been observable effects, like the decay of baryons.

There is a problem describing what happens because strictly speaking, the only observables in quantum gravity are the values of the field at infinity. One can not define the field at some point in the middle because there is quantum uncertainty in where the measurement is done. What is often done is to adopt the semi-classical approximation in which one assumes that there are a large number N of light matter fields coupled to gravity and that one can neglect the gravitational fluctuations because they are only one among N quantum loops. However, in ignoring quantum loops, one throws away unitarity. A semi-classical metric is in a mixed state already. The information loss corresponds to the classical relaxation of black holes according to the no hair theorem. One can not ask when the information gets out of a black hole because that would require the use of a semi-classical metric which has already lost the information.

In 1997, Kip Thorne and I,  bet John Preskill that information was lost in black holes. The loser or losers of the bet  were to provide the winner or winners with an encyclopedia of their own choice, from which information can be recovered with ease. I gave John an encyclopedia of baseball, but maybe I should just have given him the ashes.

\vskip.5in

\centerline{{\bf Acknowledgments}}

\vskip.3in

I am very grateful to my student, Christophe Galfard\footnote{C.Galfard@damtp.cam.ac.uk}, for help and discussions.  He is working on a proof that correlation functions decay in topologically non trivial metrics.


\begin{thebibliography}{99}


\newpage
\bibitem{Israel} W.Israel, {\it Event Horizons in Static Vacuum Space-times}, Phys. Rev. 164, 1776 (1967), and W.Israel, {\it Event Horizons in Static Electrovac Space-times}, Commun. Math. Phys. 8, 245 (1968)
\bibitem{Stephen1} S.W. Hawking, {\it Particle Creation by Black Holes}, Commun. Math. Phys. 43, 199 (1975)
\bibitem{Stephen2} S.W.Hawking, {\it Breakdown of Predictability in Gravitational Collapse}, Phys. Rev. D14, 2460 (1976),
\bibitem{ADSCFTreviews} J.M.Maldacena, {\it The Large N Limit of Superconformal Field Theory and Supergravity}, Adv. Theor. Math. Phys. 2, 231 (1998) and Int. J. Theor. Phys. 38, 1113 (1999), hep-th/9711200; S.S.Gubser, I.R.Klebanov and A.M.Polyakov, {\it Gauge Theory Correlators from Non Critical String Theory}, Phys Lett. B428, 105 (1998), hep-th/9802109  ; and a classic review: O.Aharony, J.Maldacena, S. Gubser, H.OOguri and Y.Oz, {\it Large N Field Theories, String Theory and Gravity}, Phys. Rept. 323, 183 (2000); and see E.Witten, {\it Anti-de Sitter Space and Holography}, Adv.Theor.Math.Phys.2, 253, (1998) hep-th/9802150, for the holographic point of view.

\bibitem{EQG} For an introduction, see {\it Euclidean Quantum Gravity}, Edited by G.W.Gibbons and S.W.Hawking, Singapore: World Scientific (1993)
\bibitem{Maldacena1} J.Maldacena, {\it Eternal Black Holes in Anti-de Sitter}, JHEP 0304, 21 (2003), hep-th/0106112
\bibitem{HawkingPage} S.W.Hawking and D.N.Page, {\it Thermodynamics of Black Holes in Anti-de Sitter Space}, Commun. Math. Phys. 87, 577 (1983); and E.Witten, {\it Anti-de Sitter Space, Thermal Phase Transition, And Confinement in Gauge Theories}, Adv. Theor. Math. Phys. 2, 505 (1998)
\bibitem{HartleHawking} J.B.Hartle and S.W.Hawking, {\it Path Integral Derivation of Black Hole Radiance}, Phys. Rev. D 13, 2188 (1976)
\bibitem{NoHair} B.Carter, {\it Axisymmetric Black Hole Has Only Two Degrees of Freedom}, Phys. Rev. Lett. 26, 331 (1971); S.W.Hawking, {\it Black Holes in General Relativity}, Commun. Math. Phys. 25, 152 (1972); D.C.Robinson, Phys. Rev. D 10, 458 (1974) and D.C.Robinson,  Phys. Rev. Lett 34, 905 (1975).


\end{thebibliography}
\end{document}